\documentclass{article}
\usepackage{graphicx}

\input{tcilatex}
\begin{document}

\textbf{Topological properties of fractal Julia sets related to the signs
and magnitudes of the real and reactive powers.}

H\'{e}ctor A. Tabares-Ospina\footnote{%
Facultad de Ingenier\'{\i}a, Instituci\'{o}n Universitaria Pascual Bravo,
Medell\'{\i}n, Colombia.
\par
Email: h.tabares@pascualbravo.edu.co} and John E. Candelo-Becerra\footnote{%
Facultad de Minas, Universidad Nacional de Colombia, Medell\'{\i}n, Colombia.
\par
Email: jecandelob@unal.edu.co}\vspace{0.5cm}

\bigskip \textbf{Abstract}

In AC electrical systems, the power depends on the real power (P) due to
resistive elements and the reactive power (Q) due to the inductive and
capacitive elements, which are commonly studied by using phasor and scalar
methods. Thus, this paper focuses on applying the fractal Julia sets to
observe the topological properties related to the signs and magnitudes of
the real and reactive powers consumed or supplied by an electrical circuit.
To perform this, different power combinations were used to represent the
fractal diagrams with an algorithm that considers the mathematical model of
Julia sets. The study considers three type of loads: the first study
considers the change of real power when the reactive power is fixed; the
second study deals with the change of the reactive power when the real power
is fixed; and finally, the third study contemplates that both real and
reactive powers change. Furthermore, the fractal diagrams of the power in
the four quadrants of the complex plane are studied to identify the
topological properties that each sign and magnitude represent. A qualitative
analysis of the diagrams helps to identify that the complex power loads
present some fractal graphic patterns, with respect to the signs and
magnitudes considered in the different quadrants of the complex planes. The
diagrams represented in the complex planes save a relation in the forms and
structure with other points studied, concluding that the power is related to
other figures in other quadrants. Thus, this result allows a new study of
the behavior of the power in an electrical circuit, showing a clear relation
of the different fractal diagrams that the Julia sets obtained.\vspace{0.5cm}


\newpage

\section{Introduction}

In 1975, the mathematic Benoit Mandelbrot was the responsible to define the
concept of fractal as a semi geometric element with a repetitive structure
at different scales [1], with characteristics of self-similarity as some
nature elements such as snowflakes, ferns, peacock feathers, and romanesco
broccoli. Fractal theory has been applied to various fields such as biology
[2,3], health sciences [4--8], stock markets [9], network communications
[10--12], and others. Fractal theory is one of the methods used to analyze
data and obtain relevant information in a highly complex problem. Thus, it
has been used to study the price of highly variable markets, which are not
always explainable from classical economic analyzes.

For example, in [9] the authors demonstrate that the current techniques have
some issues to explain the real market operation, and a better understanding
is achieved by using techniques as theory of chaos and fractals. In their
publication, the authors show how to apply the fractal behavior of stock
markets. Besides, the authors refer to multifractal analysis and
multifractal topology. The first describes the invariability of scaling
properties of time series and the second is a function of the H\"{o}lder
exponents that characterize the degree of irregularity of the signal and
their most significant parameters.

In [13], the authors discuss the basic principle of fractal theory and how
to use it to forecast the short-term electricity price. In the first
instance, the authors analyze the fractal characteristic of the electricity
price, confirming that price data have this property. In the second
instance, a fractal model is used to build a forecasting model, which offers
a wide application in determining the price of electricity in the markets.

Similarly, the authors of [14] demonstrate that the price of thermal coal
has multifractal features by using the concepts introduced by
Mandelbrot-Bouchaud. Hence, a quarterly fluctuation index (QFI) for thermal
power coal price is proposed to forecast the coal price caused by market
fluctuation. This study also provides a useful reference to understand the
multifractal fluctuation characteristics in other energy prices.

The fractal geometry analysis has been also applied to study the morphology
and the population growth of cities, and that the electricity demand is
related to the demography of cities. In [15] a multifractal analysis is used
to forecast the electricity demand, explaining that two fractals are found,
showing the behavior pattern of the power demand. Two concepts linked to
fractal geometry are fractal interpolation and extrapolation, related to the
resolution of a fractal-encoded image. In [16], an algorithm to forecast the
electric charge, in which fractal interpolation and extrapolation are also
involved. For the forecast data set, the average relative errors are only
2.303\% and 2.296\%. The result shows that the algorithm has advantages in
improving forecast accuracy.

In the literature, there are not papers related to the study of the signs
and magnitudes of real and reactive powers by using Mandelbrot and Julia
sets. Besides, the characteristics of the real and reactive powers are not
deeply analyzed applying the fractal geometry, concluding that these
techniques are not commonly used to study the different behaviors of the
power consumption and supply. For this reason, this work focuses on
performing a detailed analysis of the characteristics of the real and
reactive powers based on the Julia and Mandelbrot sets. Particularly in this
work, the real and reactive powers are changed to represent the different
Julia sets that allow performing observations to determine a qualitative
study. With the purpose to characterize the real and reactive power with the
Julia sets as to observe other behavior with the fractal geometry such as:
fractal graphic pattern, which requires a new interpretation to identify the
different signs. Thus, the following hypothesis is tested: the signs and
magnitudes of the real and reactive electrical powers show some related
fractal topology patterns in the Julia set, which allows to identify the
behavior of the electricity consumption and supply.

The rest of this document is organized as follows. Section 2 includes the
brief explanation of the theory of the Mandelbrot sets and Section 3
presents the theory of Julia sets. Section 4 presents the method applied in
this research and the algorithm used to identify the topological properties
of the different Julia sets of real and reactive power in electric systems.
Section 5 presents the results and discussion of the most relevant examples
of signs and magnitudes of real and reactive powers. Finally, the main
conclusions of this research work are summarized.

\subsection{Mandelbrot set}

Mandelbrot set, denoted as $M=\{$\ $c\in C/J_{c}\}$, represents the sets of
complex numbers $C$ obtained after iterating the from the initial point $%
Z_{n}$ and the selected constant $C$ as shown (\ref{Ec0}), the results form
a diagram with connected points remaining bounded in an absolute value. One
property of $M$ is that the points are connected, although in some zones of
the diagram it seems that the set is fragmented. The iteration of the
function generates a set of numbers called orbits. The results of the
iteration of those points out of the boundary set tend to infinity.

\begin{equation}
Z_{n+1}=F\left( Z_{n}\right) =Z_{n}^{2}+C  \label{Ec0}
\end{equation}

\begin{figure*}[h]\centering%
%EndExpansion
%TCIMACRO{%
%\TeXButton{Titulo}{\caption{Representation of Mandelbrot diagram}}}%
%BeginExpansion
\caption{Representation of Mandelbrot diagram}%
%EndExpansion
\label{Fig1}\includegraphics{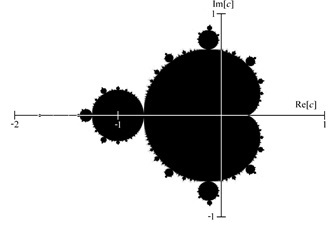}%
\begin{tabular}{l}
\FRAME{itbpF}{3.5466in}{2.3843in}{0in}{}{}{fig1.jpg}{\special{language
"Scientific Word";type "GRAPHIC";maintain-aspect-ratio TRUE;display
"USEDEF";valid_file "F";width 3.5466in;height 2.3843in;depth
0in;original-width 3.4999in;original-height 2.3436in;cropleft "0";croptop
"1";cropright "1";cropbottom "0";filename 'Fig1.jpg';file-properties
"XNPEU";}}%
\end{tabular}%
%TCIMACRO{\TeXButton{E}{\end{figure*}}}%
%BeginExpansion
\end{figure*}

From the term $C$, a successive recursion is performed with $Z_{0}=0$ as the
initial term. If this successive recursion is dimensioned, then the term $C$
belongs to the Mandelbrot set; if not, then they are excluded. Therefore,
Figure \ref{Fig1} shows the Mandelbrot set with points in the black zone
that are called the prisoners, while the points in other colors are the
escapists and they represent the velocity that they escape to infinite.

%TCIMACRO{\TeXButton{C}{\begin{figure*}[h]\centering}}%
%BeginExpansion
%EndExpansion

From this figure, the number -1 is inside of the set while the number 1 is
outside of the set. In the Mandelbrot set, the fractal is the border and the
dimension of Hausdorff is unknown. If the image is enlarged near the edge of
the set, many areas the Mandelbrot set are represented in the same form.
Besides, different types of Julia sets are distributed in different regions
of the Mandelbrot set. Whether a complex number appears with a greater value
than 2 in the 0 orbit, then the orbit tends to infinity.

The pseudocode of the algorithm that is used to represent the Mandelbrot set
is presented as follows:

\textbf{Start}

\hspace{1cm}For each point $C$ in the complex plane do:

\hspace{1cm}Fix $Z_{0}=0$

\hspace{1cm}\textbf{For} $t=1$ to $t_{\max }$ \textbf{do}:

\hspace{1cm}\hspace{1cm}Calculate $Z_{t}=Z_{t}^{2}+C$

\hspace{1cm}\hspace{1cm}\textbf{If} $|Z_{t}|>2$ \textbf{then}

\hspace{1cm}\hspace{1cm}\hspace{1cm}Break

\hspace{1cm}\hspace{1cm}\textbf{End if}

\hspace{1cm}\hspace{1cm}\textbf{If} $t<t_{\max }$ \textbf{then}

\hspace{1cm}\hspace{1cm}\hspace{1cm}Draw $C$ in white (the point does not
belong to the set)

\hspace{1cm}\hspace{1cm}\textbf{Else if} $t=t_{\max }$ \textbf{then}

\hspace{1cm}\hspace{1cm}\hspace{1cm}Draw $C$ in black (as the point does
belong to the set)

\hspace{1cm}\hspace{1cm}\textbf{End if}

\hspace{1cm}\textbf{End For}

\textbf{End\vspace{0.5cm}}

In this research, the presented algorithm has been used to obtain the
Mandelbrot set and the diagram that represent it. Some points related to the
real and reactive powers with the respective signs are studied in the
Mandelbrot set and related to those points created for the Julia sets as
explained in the following sections.

\subsection{Julia sets}

The mathematics Gast\'{o}n Julia and Pierre Fatuo, at the beginning of the
century $XX$, developed a fractal sets that are obtained by iterating
complex numbers. The Julia sets of a holomorphic function $f$ is constituted
by those points that under the iteration of $f$ have a chaotic behavior and
each point of the set forms a different set $f$ that is then denoted by $%
J(f) $. The Fatou set consists of the points that have a stable behavior
when they are iterated. The Fatou set of a holomorphic function $f$ is
denoted by $F(f)$ and it is a complement of $J(f)$. An important family of
the Julia sets is obtained from the simple quadratic functions, for example $%
Z_{n+1}=F\left( Z_{n}\right) =Z_{n}^{2}+C$, where $C$ is a complex number.
The values obtained from this function are denoted the $J_{c}$, with points
of $Z$ obtained from the parameter $C$ that belong to the Julia sets. Other
points obtained during the iteration are excluded from the Julia sets as
they tend to infinite.

For example, Figure \ref{Fig2} shows a simulation performed to obtain the
Julia sets for two complex numbers. Figure \ref{Fig2}(a) shows the Julia
sets calculated by iterating the function $Z_{n+1}=F\left( Z_{n}\right)
=Z_{n}^{2}+C$ when $C=C1=-0.33+0.57i$ and Figure \ref{Fig2}(b) shows the
Julia sets by iterating the same function when $C=C2=0.44+0.15i$. The
results show that the complex number $C1=-0.33+0.57i$ represented in Figure %
\ref{Fig2}(a) produces a Julia set of points in black and other points with
different colors tend to the infinity according to the number of iterations
they require to obtain the result. However, the complex number $%
C2=0.44+0.15i $ represented in Figure \ref{Fig2}(b) produces a Julia set
with non-connected points.

\includegraphics{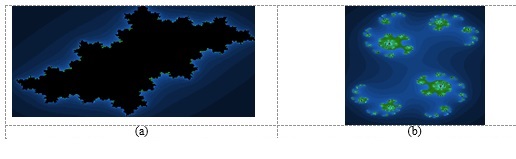}

%TCIMACRO{\TeXButton{C}{\begin{figure*}[h]\centering}}%
%BeginExpansion
\begin{figure*}[h]\centering%
%EndExpansion
%TCIMACRO{\TeXButton{Titulo}{\caption{Julia sets for two complex numbers}}}%
%BeginExpansion
\caption{Julia sets for two complex numbers}%
%EndExpansion
\label{Fig2}%
\begin{tabular}{ll}
&  \\ 
(a) $C1=-0.33+0.57i$ & (b) $C2=0.44+0.15i$%
\end{tabular}%
%TCIMACRO{\TeXButton{E}{\end{figure*}}}%
%BeginExpansion
\end{figure*}%
%EndExpansion

An important relation between the Mandelbrot and the Julia sets is given
when the point $C$ belongs to the Mandelbrot set, then the Julia set $%
J(f_{c})$ obtain a series of points that are connected. On the other side,
when the point does not belong to the Mandelbrot set, then the Julia set $%
J(f_{c})$ is formed by non-connected points. Therefore, there are two main
classes of Julia sets: points that form a unique piece and points that
represent many pieces in the complex plane.

One property of the Mandelbrot set is that the different types of Julia sets
are distributed in different regions of the set $M$. For all the above, it
is concluded that in Figure \ref{Fig2}, $C1$ is in the set of $M$, and $C2$
is not in the set. In general, any point within the $M$ cardioid or its
boundary, the Julia set of the $J(f_{c})$ has points that are connected. The
most interesting $C$ values are those near the border of the Mandelbrot set,
because it is there when the set $J(f_{c})$ passes, suddenly, from connected
points to non-connected points.

\section{Procedure implemented in the research}

As the electrical power can be expressed as a complex expression given by $%
S=P+jQ$, the combination values with the respective signs and magnitudes of
the real and reactive powers are studied through the representation of the
Julia sets, obtaining fractal geometric patterns that allow a qualitative
analysis. To validate the hypothesis, the procedure of this research was
developed to evaluate certain properties of the electrical power by using
the fractal geometry.

The first part of the procedure starts with the analysis of the real and
reactive power in the complex plane of the electric power. The following
cases are considered in the study:

\begin{enumerate}
\item[a)] The real power $P>0$ and the reactive power $Q>0$ corresponds to
the first complex plane.

\item[b)] The real power $P<0$ and the reactive power $Q>0$ corresponds to
the second complex plane.

\item[c)] The real power $P<0$ and the reactive power $Q<0$ corresponds to
the third complex plane.

\item[d)] The real power $P>0$ and the reactive power $Q<0$ corresponds to
the fourth complex plane.
\end{enumerate}

In the second part, an algorithm was programmed to perform the numerical
experiments on the Mandelbrot and Julia sets by using the real and reactive
power with different power factors. To test the system, the most
representative Julia sets were executed.

As the $M$ set has infinite values in the quotation, the third part of this
work dedicates on determining two quotation points. The limits of $M$ in the
positive axis $(x,y)$ are represented in the diagrams and the values of $P$
and $Q$ are scale in the set of $M$ with values $C_{x}=0.25$ and $C_{y}=0.63$%
.

In order to validate the work hypothesis, the fourth part consisted of
operating the program to find a pattern from the fractal geometries,
depending on the different signs of the real and reactive powers. In the
fifth part, Julia sets were classified, according to their symmetry and
degree of irregularity in their borders, selecting the most interesting by
their degree of dispersion. Finally, the Julia sets related to each complex
plane are found and the conclusions are obtained.

\newpage 

\section{Results and analysis}

The following figures present the diagrams created with the Julia sets
generated from the real and reactive components of the electrical power.
These generated fractal geometries do not give a quantitative information of
the fact under study; however, the study of the fractals can be performed by
a qualitative analysis, whose interpretation is based on the experience.

\subsection{Julia sets for the real power}

In order to simplify the analysis, the study is performed first for the
change of real power with the power factor equal to $1$ or the reactive
power equals to $zero$. This study relates to a resistive circuit where the
voltage $(V)$ is in phase with the current. Thus, Figure \ref{Fig3} presents
the transformation of the geometric fractal patterns according to the change
of real power. These figures are plotted in the complex plane, in which the $%
x$ axis corresponds to the real component $(Re)$ and the $y$ axis
corresponds to the imaginary component $(Img)$. Figure \ref{Fig3}(a) shows
the fractal when real and reactive powers are equal to $cero$, which obtains
as a result a circle with radius equal to $1$. Figure \ref{Fig3}(b) presents
the fractal when $P=D_{x}/2$, showing a transformation of the closed fractal
curve and evidencing contraction in its symmetry axes. Figure \ref{Fig3}(c)
shows a fractal when $P=D_{x}$, showing that the Julia set is close to the
limit of the set $M$ or to separate the points. Figure \ref{Fig3}(d) shows
that when $P>D_{x}$ or $P$ exceeds the limit of the set $M$, the Julia set
presents several points that are not connected or explode in a point cloud.

\includegraphics{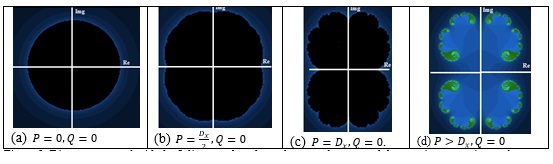}

%TCIMACRO{\TeXButton{C}{\begin{figure*}[h]\centering}}%
%BeginExpansion
\begin{figure*}[h]\centering%
%EndExpansion
%TCIMACRO{%
%\TeXButton{Titulo}{\caption{Diagrams created with the Julia sets when the real power changes and the reactive power is equal to cero}}}%
%BeginExpansion
\caption{Diagrams created with the Julia sets when the real power changes and the reactive power is equal to cero}%
%EndExpansion
\label{Fig3}%
\begin{tabular}{llll}
&  &  &  \\ 
(a) $P=0,Q=0$ & (b) $P=\frac{D_{x}}{2},Q=0$ & (c) $P=D_{x},Q=0$ & (d) $%
P>D_{x},Q=0$%
\end{tabular}%
%TCIMACRO{\TeXButton{E}{\end{figure*}}}%
%BeginExpansion
\end{figure*}%
%EndExpansion

Therefore, the Julia sets form a simple closed curve that transforms to a
fractal form with the following relations: the upper and lower half-planes
are reflections of each other, while the right and left half planes are
reflections of each other. Besides, the points of the upper and lower
half-planes are symmetrical with respect to the real axis of the complex
plane, while the points of the right and left half planes are symmetrical
with respect to the imaginary axis of the complex plane.

In conclusion, reflective symmetry is presented with respect to the origin
point and it is confirmed that there exists a fractal geometric pattern as a
function of $P$. The relation found in this research is because the origin
Mandelbrot set is symmetric with respect to the real axis of the complex
plane. With respect to the fractal curve in the limits of the set, it
reveals the approximation that has the magnitude of the real power with
respect to the limit of the set of $M$. If the real power exceeds the limit
of $M$ on the x axis, the Julia set is not connected and presents several
points as exploding.

\newpage 

\subsection{Julia sets for the reactive power}

Now, the fractal geometry obtained as a function of reactive power is
studied. This case represents a motor or generator without load, which
considers only reactive power for the magnetization of its coils, which is
also evidenced by a low power factor. Thus, Figure \ref{Fig4} shows the
transformation of the fractal geometry as a function of the change of the
reactive power $Q$. Figure \ref{Fig4}(a) shows the fractal geometry when $P$
and $Q$ are equal to $zero$, which represents is a circle of radius equal to 
$1$. Figure \ref{Fig4}(b) shows the fractal geometry when $D_{y}/2$, which
is transformed into a closed fractal curve. Figure \ref{Fig4}(c) shows the
fractal geometry when $Q=D_{y}$, presenting the Julia set next to its
separation because $Q$ is above the limit of $M$. Figure \ref{Fig4}(d) shows
the fractal geometry when $Q>D_{y}$, which exceeds the limit of the set of $M
$, and the points of the Julia sets are not connected, with separated zones.

\includegraphics{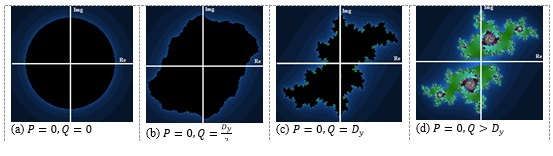}

%TCIMACRO{\TeXButton{C}{\begin{figure*}[h]\centering}}%
%BeginExpansion
\begin{figure*}[h]\centering%
%EndExpansion
%TCIMACRO{%
%\TeXButton{Titulo}{\caption{Diagrams obtained with the Julia sets in function of the reactive power Q}}}%
%BeginExpansion
\caption{Diagrams obtained with the Julia sets in function of the reactive power Q}%
%EndExpansion
\label{Fig4}%
\begin{tabular}{llll}
&  &  &  \\ 
(a) $P=0,Q=0$ & (b) $P=0,Q=\frac{D_{y}}{2}$ & (c) $P=0,Q=D_{y}$ & (d) $%
P=0,Q>D_{y}$%
\end{tabular}%
%TCIMACRO{\TeXButton{E}{\end{figure*}}}%
%BeginExpansion
\end{figure*}%
%EndExpansion

In this case, there are different fractal patterns when the reactive power
is studied, showing the inverse reflective symmetry with respect to the
origin. In Figures \ref{Fig4}(a), (b), and (c), the Julia set is a simple
closed curve that becomes a fractal curve. However, unlike the relationships
shown in \ref{Fig3}(a), (b) and (c), in this case the half-planes keep
inverted reflections of one another. The relationship between the quadrants
is the following: the upper and lower half-planes are reflections inverted
one from the other, while the right and left half planes are reflections
inverted one from the other. The points of the upper and lower half planes
are symmetrical inverted with respect to the real axis of the complex plane,
while the points of the right and left half planes are symmetrical inverted
with respect to the imaginary axis of the complex plane.

In conclusion, it is confirmed that there exists another fractal geometric
pattern for the reactive power $Q$, which show the inverted reflective
symmetry with respect to the origin point is presented. The fractal geometry
in the limits reveals the approximation that the magnitude of the reactive
power has with respect to the limit of the set of $M$. If the reactive power
exceeds the limit of $M$ on the y axis, the Julia sets becomes not
connected, exploding in point cloud.

\newpage 

\subsection{Julia sets of the real and reactive power values}

Next, the fractal diagrams obtained from the changes of real and reactive
powers are presented. Therefore, the following figures show fractals for the
four quadrants. To perform the tests and show the results clearly, all
values were considered with the values $P\pm 0.22$, $Q\pm 0.22$.

\subsubsection{Analysis on the first quadrant}

Figure 5 shows the fractal diagrams related to the Mandelbrot and Julia sets
when the real and reactive power is located in the first quadrant of the
complex plane. Figure \ref{Fig5}(a) shows the real and reactive powers with
an inductive element. Figure \ref{Fig5}(b) shows the Mandelbrot diagram
created by the algorithm that iterates the complex number created by the
real and reactive power $(P=0.22,Q=0.22)$. Besides, a point in yellow is
marked in this figure to adjust the real and reactive power to the diagram.
Figure \ref{Fig5}(c) shows the Julia set that is created from the point
selected from the Mandelbrot diagram and related to the real and reactive
power studied. In general, when $P$ and $Q$ are positive and located within
the set $M$, then the equivalent set $J(f_{c})$ is connected and its graph
is a closed curve with inverted reflective symmetry with respect to the
origin point, and with dendrites oriented in a clockwise direction.

\includegraphics{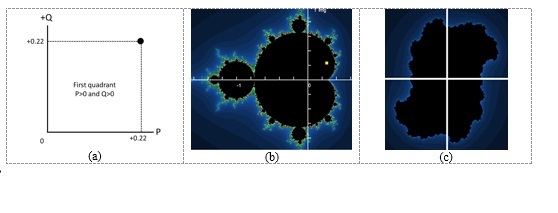}

%TCIMACRO{\TeXButton{C}{\begin{figure*}[h]\centering}}%
%BeginExpansion
\begin{figure*}[h]\centering%
%EndExpansion
%TCIMACRO{%
%\TeXButton{Titulo}{\caption{Mandelbrot and Julia sets for a real and reactive power in the first complex plane}}}%
%BeginExpansion
\caption{Mandelbrot and Julia sets for a real and reactive power in the first complex plane}%
%EndExpansion
\label{Fig5}%
\begin{tabular}{lll}
&  &  \\ 
(a) first quadrant & (b) Mandelbrot diagram & (c) Julia diagram%
\end{tabular}%
%TCIMACRO{\TeXButton{E}{\end{figure*}}}%
%BeginExpansion
\end{figure*}%
%EndExpansion

\newpage 

\subsubsection{Analysis on the fourth quadrant}

Figure 6 shows the fractal diagrams related to the Mandelbrot and Julia sets
when the power is at the fourth quadrant of the complex plane. Figure \ref%
{Fig6}(a) shows the real and reactive power with a capacitive element ($%
P=0.22,Q=-0.22$). Figure \ref{Fig6}(b) shows the Mandelbrot diagram created
by the algorithm that iterates the complex number created by the real and
reactive power. Besides, a point in yellow is marked in this figure to
adjust the real and reactive power to the diagram. Figure \ref{Fig6}(c)
shows the Julia set that is created from the point selected from the
Mandelbrot diagram and related to the real and reactive power studied.

\includegraphics{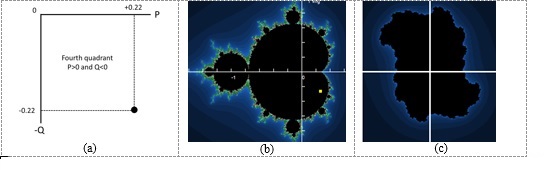}

%TCIMACRO{\TeXButton{C}{\begin{figure*}[h]\centering}}%
%BeginExpansion
\begin{figure*}[h]\centering%
%EndExpansion
%TCIMACRO{%
%\TeXButton{Titulo}{\caption{Mandelbrot and Julia sets for a real and reactive power in the fourth complex plane}}}%
%BeginExpansion
\caption{Mandelbrot and Julia sets for a real and reactive power in the fourth complex plane}%
%EndExpansion
\label{Fig6}%
\begin{tabular}{lll}
&  &  \\ 
(a) fourth & (b) Mandelbrot diagram & (c) Julia diagram%
\end{tabular}%
%TCIMACRO{\TeXButton{E}{\end{figure*}}}%
%BeginExpansion
\end{figure*}%
%EndExpansion

Figures \ref{Fig5} and \ref{Fig6} show that the forms have the same relation
but they are symmetrically inverted with respect to the origin and the
equivalent set $J(f_{c})$ is connected and with closed curve, with dendrites
oriented anti-clockwise.

\newpage 

\subsubsection{Analysis on the second quadrant}

Figure 7 shows the fractal diagrams related to the Mandelbrot and Julia sets
when the power is at the second quadrant of the complex plane. Figure \ref%
{Fig7}(a) shows the real power supplied to the circuit with a capacitive
reactive power $(P=-0.22,Q=0.22)$. Figure \ref{Fig7}(b) shows the Mandelbrot
diagram created by the algorithm that iterates the complex number created
from the power of the electric circuit. Besides, a point in yellow is marked
in this figure to adjust the real and reactive power to the diagram. Figure %
\ref{Fig7}(c) shows the Julia set that is created from the point selected
from the Mandelbrot diagram and related to the real and reactive powers
studied. When this values are in $M$, the equivalent set $J(f_{c})$ is
connected and the figure forms a closed fractal different to the other two
cases, with apparent clockwise direction with respect to the real axis of
the complex plane.

\includegraphics{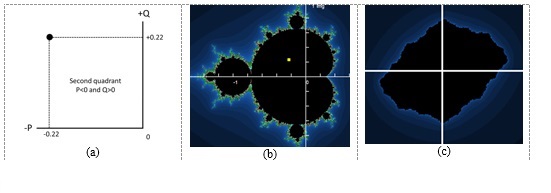}

%TCIMACRO{\TeXButton{C}{\begin{figure*}[h]\centering}}%
%BeginExpansion
\begin{figure*}[h]\centering%
%EndExpansion
%TCIMACRO{%
%\TeXButton{Titulo}{\caption{Mandelbrot and Julia sets for a real and reactive power in the second complex plane}}}%
%BeginExpansion
\caption{Mandelbrot and Julia sets for a real and reactive power in the second complex plane}%
%EndExpansion
\label{Fig7}%
\begin{tabular}{lll}
&  &  \\ 
(a) second quadrant & (b) Mandelbrot diagram & (c) Julia diagram%
\end{tabular}%
%TCIMACRO{\TeXButton{E}{\end{figure*}}}%
%BeginExpansion
\end{figure*}%
%EndExpansion

\newpage 

\subsubsection{Analysis on the third quadrant}

Figure \ref{Fig8} shows a new fractal diagrams related to the Mandelbrot and
Julia sets when the power is at the third quadrant of the complex plane.
Figure \ref{Fig8}(a) shows the real power supplied to the circuit with an
inductive reactive power $(P=-0.22,Q=-0.22)$. Figure \ref{Fig8}(b) shows the
Mandelbrot diagram created by the algorithm that iterates the complex number
created from the power of the electric circuit. Besides, a point in yellow
is marked in this figure to adjust the real and reactive power to the
diagram. Figure \ref{Fig8}(c) shows the diagram created with the Julia set
from the point selected in the Mandelbrot diagram and related to the real
and reactive power studied. Figure \ref{Fig8} presents a fractal geometry of
Julia set with inverted symmetry with respect to Figure \ref{Fig7}. When the
power is in the third quadrant and inside the main bulb of the set of $M$,
the equivalent set $J(f_{c})$ is connected and its graph is a closed curve,
which appears in a clockwise rotation with respect to the real axis of the
complex plane.

\includegraphics{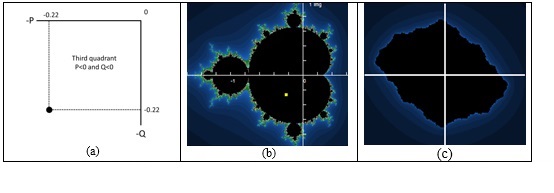}

%TCIMACRO{\TeXButton{C}{\begin{figure*}[h]\centering}}%
%BeginExpansion
\begin{figure*}[h]\centering%
%EndExpansion
%TCIMACRO{%
%\TeXButton{Titulo}{\caption{Mandelbrot and Julia sets for a power that is in the third complex plane}}}%
%BeginExpansion
\caption{Mandelbrot and Julia sets for a power that is in the third complex plane}%
%EndExpansion
\label{Fig8}%
\begin{tabular}{lll}
&  &  \\ 
(a) third quadrant & (b) Mandelbrot diagram & (c) Julia diagram%
\end{tabular}%
%TCIMACRO{\TeXButton{E}{\end{figure*}}}%
%BeginExpansion
\end{figure*}%
%EndExpansion

Figures \ref{Fig7} and \ref{Fig8} showed that the forms have the same
relation, but they are symmetrically inverted with respect to the origin.
Again, a qualitative interpretation is that the capacitive and inductive
electric power corresponds to two types of electrical phenomena contrary to
each other. These two figures showed the behavior of the power with fractal
diagrams: the first figure represented the reactive power related to
capacitive elements as the current leads the voltage; and the second figure
represented the reactive power related to inductive elements as the current
lags the voltage.

\newpage 

\section{Conclusions}

This paper used the fractal Julia sets to observe the topological properties
related to the signs and magnitudes of the real and reactive powers consumed
or supplied by an electrical circuit. The method consisted in considering
different power combinations to represent the fractal diagrams with an
algorithm that iterates the mathematical expressions of the Julia sets. The
study considered the change of real power when the reactive power is fixed,
the change of the reactive power when the real power is fixed, and the
change of both real and reactive powers. Furthermore, the fractal diagrams
of the power in the four quadrants of the complex plane were studied to
identify the topological properties that each sign and magnitude represent.

The study of the fractal diagrams helped to identify that the real and
reactive powers represents clear fractal patterns with the Julia sets. When
the reactive power is greater than zero and the real power changes, the
Julia set presents an inverted symmetry with respect to the origin point.
Besides, when the real and reactive powers are equal to zero, the fractal
diagram is a circle with a radius of one in the complex plane. As a function
of real power, two Julia sets are generated with different fractal patterns
that characterize the signs and magnitudes. The characteristics of the Julia
sets depend on the values of the initial point $Z_{n}$. Additionally, these
are groups with a chaotic behavior, because their fractal curve cannot be
predicted. When numerical analysis is repeated for different real and
reactive powers, they produce different fractal curves that are
symmetrically inverted with respect to the origin, as explained in this work.

\section{Acknowledgement}

This work was supported by the Agencia de Educaci\'{o}n Superior de Medell%
\'{\i}n (Sapiencia), under the specific agreement celebrated with the
Instituci\'{o}n Universitaria Pascual Bravo. The project is part of the
Energy System Doctorate Program and the Department of Electrical Energy and
Automation of the Universidad Nacional de Colombia, Sede Medell\'{\i}n,
Facultad de Minas.

\section{References}

[1]\qquad Barnsley, M. F., Devaney, R. L., Mandelbrot, B. B., Peitgen,
H.-O., Saupe, D., and Voss, R. F., 1988, The Science of Fractal Images,
Springer New York, New York, NY.

[2]\qquad Strogatz, S. H., 2018, Nonlinear Dynamics and Chaos, CRC Press.

[3]\qquad Losa, G. A., 2012, \textquotedblleft Fractals and Their
Contribution to Biology and Medicine,\textquotedblright\ Medicographia, 34,
pp. 365--374.

[4]\qquad Garcia, T. A., Tamura Ozaki, G. A., Castoldi, R. C., Koike, T. E.,
Trindade Camargo, R. C., and Silva Camargo Filho, J. C., 2018,
\textquotedblleft Fractal Dimension in the Evaluation of Different
Treatments of Muscular Injury in Rats,\textquotedblright\ Tissue Cell, 54,
pp. 120--126.

[5]\qquad Rodr\'{\i}guez V, J. O., Prieto B, S. E., Correa H, S. C.,
Soracipa M, M. Y., Mendez P, L. R., Bernal C, H. J., Hoyos O, N. C., Valero,
L. P., Velasco R, A., and Bermudez, E., 2016, \textquotedblleft Nueva
Metodolog\'{\i}a de Evaluaci\'{o}n Del Holter Basada En Los Sistemas Din\'{a}%
micos y La Geometr\'{\i}a Fractal: Confirmaci\'{o}n de Su Aplicabilidad a
Nivel Cl\'{\i}nico,\textquotedblright\ Rev. la Univ. Ind. Santander. Salud,
48(1), pp. 27--36.

[6]\qquad Popovic, N., Radunovic, M., Badnjar, J., and Popovic, T., 2018,
\textquotedblleft Fractal Dimension and Lacunarity Analysis of Retinal
Microvascular Morphology in Hypertension and Diabetes,\textquotedblright\
Microvasc. Res., 118, pp. 36--43.

[7]\qquad Hern\'{a}ndez Vel\'{a}zquez, J. de D., Mej\'{\i}a-Rosales, S., and
Gama Goicochea, A., 2018, \textquotedblleft Fractal Properties of
Biophysical Models of Pericellular Brushes Can Be Used to Differentiate
between Cancerous and Normal Cervical Epithelial Cells,\textquotedblright\
Colloids Surfaces B Biointerfaces, 170, pp. 572--577.

[8]\qquad Moon, P., Muday, J., Raynor, S., Schirillo, J., Boydston, C.,
Fairbanks, M. S., and Taylor, R. P., 2014, \textquotedblleft Fractal Images
Induce Fractal Pupil Dilations and Constrictions,\textquotedblright\ Int. J.
Psychophysiol., 93(3), pp. 316--321.

[9]\qquad Mandelbrot, B. B., and Hudson, R. L., 2004, The (Mis) Behaviour of
Markets: A Fractal View of Risk, Ruin and Reward, Profile Books, London.

[10]\qquad Kumar, R., and Chaubey, P. N., 2012, \textquotedblleft On the
Design of Tree-Type Ultra Wideband Fractal Antenna for DS-CDMA
System,\textquotedblright\ J. Microwaves, Optoelectron. Electromagn. Appl.,
11(1), pp. 107--121.

[11]\qquad Ma, Y.-J., and Zhai, M.-Y., 2018, \textquotedblleft Fractal and
Multi-Fractal Features of the Broadband Power Line Communication
Signals,\textquotedblright\ Comput. Electr. Eng.

[12]\qquad Ye, D., Dai, M., Sun, Y., and Su, W., 2017, \textquotedblleft
Average Weighted Receiving Time on the Non-Homogeneous Double-Weighted
Fractal Networks,\textquotedblright\ Phys. A Stat. Mech. its Appl., 473, pp.
390--402.

[13]\qquad Cui, H., and Yang, L., 2009, \textquotedblleft Short-Term
Electricity Price Forecast Based on Improved Fractal
Theory,\textquotedblright\ 2009 International Conference on Computer
Engineering and Technology, IEEE, pp. 347--351.

[14]\qquad Zhao, Z., Zhu, J., and Xia, B., 2016, \textquotedblleft
Multi-Fractal Fluctuation Features of Thermal Power Coal Price in
China,\textquotedblright\ Energy, 117, pp. 10--18.

[15]\qquad Salv\'{o}, G., and Piacquadio, M. N., 2017, \textquotedblleft
Multifractal Analysis of Electricity Demand as a Tool for Spatial
Forecasting,\textquotedblright\ Energy Sustain. Dev., 38, pp. 67--76.

[16]\qquad Zhai, M.-Y., 2015, \textquotedblleft A New Method for Short-Term
Load Forecasting Based on Fractal Interpretation and Wavelet
Analysis,\textquotedblright\ Int. J. Electr. Power Energy Syst., 69, pp.
241--245.

\newpage 

\section{Bibliography of authors}

%TCIMACRO{\TeXButton{C}{\begin{figure*}[h]\centering}}%
%BeginExpansion
\begin{figure*}[h]\centering%
%EndExpansion
%TCIMACRO{\TeXButton{Titulo}{\caption{Authors}}}%
%BeginExpansion
\caption{Authors}%
%EndExpansion
\label{Fig10}%
\begin{tabular}{ll}
\includegraphics{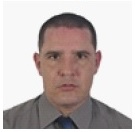} & \includegraphics{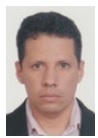} \\ 
First Author's & Second author's%
\end{tabular}%
%TCIMACRO{\TeXButton{E}{\end{figure*}}}%
%BeginExpansion
\end{figure*}%
%EndExpansion

Firts Author's: H\'{e}ctor A. Tabares-Ospina: received his Bs. degree in
Electrical Engineering in 1997 and his Master in Systems Engineering in 2005
from Universidad Nacional de Colombia. He is now studing doctoral studies.
He is an Assistant Professor of Instituci\'{o}n Universitaria Pascual Bravo.
His research interests include: Fractal geometry, artificial intelligence,
operation and control of power systems; and smart grids. He is a Junior
Researcher in Colciencias and member of the Research Group - GIIEN, at
Instituci\'{o}n Universitaria Pascual Bravo.
https://orcid.org/0000-0003-2841-6262\\*[0pt]

Second Author's:John E. Candelo-Becerra: received his Bs. degree in
Electrical Engineering in 2002 and his PhD in Engineering with emphasis in
Electrical Engineering in 2009 from Universidad del Valle, Cali - Colombia.
His employment experiences include the Empresa de Energ\'{\i}a del Pac\'{\i}%
fico EPSA, Universidad del Norte, and Universidad Nacional de Colombia -
Sede Medell\'{\i}n. He is now an Assistant Professor of the Universidad
Nacional de Colombia - Sede Medell\'{\i}n, Colombia. His research interests
include: engineering education; planning, operation and control of power
systems; artificial intelligence; and smart grids. He is a Senior Researcher
in Colciencias and member of the Applied Technologies Research Group - GITA,
at the Universidad Nacional de Colombia.
https://orcid.org/0000-0002-9784-9494.

\end{document}